\documentclass{aa}
\usepackage{graphicx}
\usepackage{epsfig}
\usepackage{natbib}
\usepackage{amssymb}
\newcommand{\De}{\ensuremath{{\rm D}}}

\newcommand{\He}{\ensuremath{\null^4{\rm He}}}

\newcommand{\Li}{\ensuremath{\null^7{\rm Li}}}

\begin{document}

\title{Time variation of the fine structure constant in the early
  universe and the Bekenstein model} 

\author{M. E. Mosquera$^{1}$ \thanks{fellow of CONICET}  \and C.
  G. Sc\'{o}ccola$^{1,2}$ \thanks{fellow of CONICET} \and
  S. J. Landau$^{3}$ \thanks{member of  the Carrera del Investigador
    Cient\'{\i}fico y Tecnol\'ogico, CONICET} \and  H. Vucetich$^{1}$ } 

\institute{Facultad de Ciencias Astron\'{o}micas y Geof\'{\i}sicas,
  Universidad Nacional de La Plata, Paseo del Bosque, cp 1900 La
  Plata, Argentina\\ 
\and 
Instituto de Astrof{\'\i}sica La Plata \\
\and
Departamento de F{\'\i}sica, FCEyN, Universidad de Buenos Aires,
  Ciudad Universitaria - Pab. 1, 1428 Buenos Aires, Argentina\\ 
\email{mmosquera, cscoccola, vucetich@fcaglp.unlp.edu.ar,
  slandau@df.uba.ar}} 

\date{Received; accepted}

\abstract {} {We calculate bounds on the variation of the fine
structure constant at the time of primordial nucleosynthesis and at
the time of neutral hydrogen formation. We use these bounds and other
bounds from the late universe to test Bekenstein model.}  {We modify
the Kawano code, CAMB and CosmoMC in order to include the possible
variation of the fine structure constant.  We use observational
primordial abundances of $\De$, $\He$ and $\Li$, recent data
from the Cosmic Microwave Background and the 2dFGRS power spectrum, to
obtain bounds on the variation of $\alpha$. We calculate a piecewise
solution to the scalar field equation of Bekenstein model in two
different regimes; i) matter and radiation, ii) matter and
cosmological constant. We match both solutions with appropriate boundary
conditions. We perform a statistical analysis using the bounds
obtained from the early universe and other bounds from the late
universe to constrain the free parameters of the model.}  {Results are
consistent with no variation of $\alpha$ in the early universe. Limits
on $\alpha$ are inconsistent with the scale length of the theory $l$
being larger than Planck scale.}  {In order to fit all observational
and experimental data, the assumption $l>L_p$ implied in Bekenstein's
model has to be relaxed.}

\authorrunning{Mosquera et al.}

\titlerunning{Variation of $\alpha$ and the Bekenstein model}

\maketitle

\section{Introduction}
\label{Intro}

Time variation of fundamental constants has been an active field of
theoretical and experimental research since the large number
hypothesis (LNH) was proposed by \citet{Dirac37}. The great predictive
power of the LNH induced a large number of research papers and
suggested new sources of variation. Among them, the attempt to unify
all fundamental interactions resulted in the development of
multidimensional theories, like string derived field theories
\citep{Wu86,Maeda88,Barr88,DP94,DPV2002a,DPV2002b}, related
brane-world theories \citep{Youm2001a,Youm2001b,branes03a,branes03b},
and Kaluza-Klein theories \citep{Kaluza,Klein,Weinberg83,GT85,OW97},
where the gauge coupling constants may vary over cosmological time
scales.

Following a different path of research, \citet{Bekenstein82} proposed
a theoretical framework to study the fine structure constant
variability based on general assumptions: covariance, gauge
invariance, causality and time-reversal invariance of
electromagnetism, as well as the idea that the Planck-Wheeler length
$\left( 10^{-33}{\rm cm} \right)$ is the shortest scale allowable in
any theory. It is well known, that bounds from the weak equivalence
principle require $l<L_p$. However, in this paper we are going to
analyze data from cosmological time-scales rather than planetary
scales, the latter ones being relevant to probe the validity of weak
equivalence principle. The model was improved by \citet{BSM02} using the main  
assumption that cold dark matter has magnetic fields dominating its
electric fields. Moreover, a super symmetric generalization of this
model was performed by \citet{OP02}, allowing additional couplings
between the scalar field and both a dark matter candidate and the
cosmological constant.  The model was also generalized in order to
study the time variation of the strong coupling constant by
\citet{CLV01}.

Different versions of the theories mentioned above predict different
time behavior of the gauge coupling constants. Thus, bounds obtained
from astronomical and geophysical data are an important tool to test
the validity of these theories. In unifying schemes like the ones
described above, the variation of each gauge coupling constant is
related to the others. In this paper, we limit ourselves to study the
variation of the fine structure constant ($\alpha$).

The experimental research can be grouped in astronomical and local
methods. The latter ones include geophysical methods such as the
natural nuclear reactor that operated about $1.8 \times 10^9$ years
ago in Oklo, Gabon, the analysis of natural long-lived $\beta$
decayers in geological minerals and meteorites and laboratory
measurements such as comparisons of rates between clocks with
different atomic number. The astronomical methods are based mainly in
the analysis of spectra from high redshift quasar absorption systems.
Although most of the previously mentioned experimental data gave null
results, evidence of time variation of the fine structure constant was
reported recently from high redshift quasar absorption systems
\citep{Webb99,Webb01,Murphy01a,Murphy01b,Murphy03b,Levshakov07}. However,
other recent independent analysis of similar data
\citep{MVB04,QRL04,Bahcall04,Srianand04} found no variation.

Bounds on the variation of $\alpha$ in the early universe can be
obtained using data from the Cosmic Microwave Background (CMB)
radiation and from the abundances of light elements.  Although these
bounds are not as stringent as the mentioned above, they are important
because they refer to a different cosmological epoch. Finally, 
other bounds at redshift greater than $30$ could be obtained from the 21 cm
signal once it could be measured \citep{KW07}. In this paper, we
perform a careful analysis of the time variation of $\alpha$ in the
early universe. First, we use available abundances of $\De$,
$\He$ and $\Li$ and the latest data from the CMB to put bounds
on the variation of $\alpha$ in the early universe without assuming
any theoretical model. Then, we use these bounds and others from
astronomical and geophysical data, to test Bekenstein theory. 

In section \ref{nucleo} we use the abundances of the light elements to
put bounds on $\frac{\Delta \alpha}{\alpha_0}$, where $\alpha_0$ is
the actual value of $\alpha$, allowing the baryon to photon density
$\eta_B$ to vary. We also calculate the time variation of $\alpha$
keeping $\eta_B$ fixed to the WMAP estimation. In section \ref{cmb} we
use the three year WMAP, other CMB experiments and the power spectrum
of the 2dFGRS to put bounds on the variation of $\alpha$ during
recombination, allowing also other cosmological parameters to vary. In
sections \ref{quasars}, \ref{geo} and \ref{lab} we describe the
astronomical and local data from the late universe.  In section
\ref{modelo} we describe Bekenstein model for $\alpha$ variations and
obtain solutions for the early and late universe. In section
\ref{resultados} we show our results. Finally, in section
\ref{conclusiones} we discuss the results and summarize our
conclusions. 

\section{Bounds from BBN}
\label{nucleo}

Big Bang Nucleosynthesis (BBN) is one of the most important tools to
study the early universe. The model is simple and has only one free
parameter, the baryon to photon ratio $\eta_B$, which can be
determined by comparison between theoretical calculations and
observations of the abundances of light elements. On the other hand,
data from the Cosmic Microwave Background (CMB) provide an independent
method for determining $\eta_B$ \citep{wmapest,wmap3,Sanchez06}.
Considering the baryon density from WMAP results, the predicted
abundances are highly consistent with the observed $\De$ but not with
all $\He$ and $\Li$.  Such discrepancy is usually ascribed to non
reported systematic errors in the observations of $\He$ and
$\Li$. However, if the systematic errors of $\He$ and $\Li$ are
correctly estimated, we may have insight into new physics beyond the
minimal BBN model. \citet{dmitriev04} consider the variation of the
deuterium binding energy in order to solve the discrepancy between
$\De$, $\He$ and $\Li$ abundances.

In this section we focus on the possibility that the fine structure
constant may be different from its actual value during BBN. The
dependence of the primordial abundances on the fine structure constant
has been evaluated by \citet{Iguri99} and improved by
\citet{Nollet}. Semi-analytical analysis have been performed by some
of us in earlier works \citep{LMV06,Chamoun07}. \citet{ichi02}
study the effects of variation of fundamental constants on BBN in the
context of a dilaton superstring model. In a following work, they study
the primordial abundances of light elements when the fine structure
constant and the cosmic expansion rate take non-standard values
\citep{Ichi04}. \citet{muller04} calculate the primordial abundances
as a function of the Planck mass, fine structure constant, 
Higgs vacuum expectation value, electron mass, nucleon decay time,
deuterium binding energy and neutron-proton mass difference and study
the dependence of the last three quantities as functions of the
fundamental coupling and masses. \citet{coc07} set constraints on the
variation of the neutron lifetime and neutron-proton mass difference
using the primordial abundance of $^4{\rm He}$. Then, they translate
these constraints into bounds on the time variation of the Yukawa
couplings and the fine structure constant. \citet{cyburt05} study the
number of relativistic species at the time of BBN and the variation of
fundamental constants $\alpha$ and $G_N$ and set bounds in these
quantities using the primordial abundances and the results of WMAP for
$\eta_B$.

In this work, we modify numerical code of Kawano \citep{Kawano88,Kawano92} in order to
allow $\alpha$ to vary. In addition to the dependences on
$\alpha$ discussed by other authors, we also include the dependence of
the light nuclei masses on $\alpha$ \citep{LMV06}. The code was
also updated with the reactions rates reported by \citet{Iguri99}.

We consider available observational data on $\De$, $\He$ and
$\Li$. For $\De$ we consider the values reported by
\citet{pettini,omeara,kirkman,burles1,burles2,Crighton04,omeara06,oliveira06}.

For $\Li$ we consider the results from
\citet{ryan,bonifacio1,bonifacio2,bonifacio3,Asplund05,BNS05,bonifacio07}.

The $\He$ available observations can be summarized in the results
reported by \citet{PL07,izotov07,olive07}. The reported
values of $\He$ depend on the adopted set of $\rm{He} I $
emissivities. In fact, \citet{izotov07} report two values, one
calculated with old atomic data \citep{Benjamin02} and the other with
new atomic data \citep{Porter05} while \citet{PL07} used the new values.
We consider the results calculated using new atomic data. \citet{olive07}
reanalyze the values of \citet{izotov1,peimbert1} for the primordial
abundance of $^4{\rm He}$. They examine some sources of systematics
uncertainties and conclude that the observational determination of
primordial helium abundance is limited by systematics errors. We
consider the data of \citet{PL07,izotov07} in our analysis.  In table \ref{standard-abundances} we show the theoretical predictions of the abundances in the standard model (without variation of $\alpha$) with $\eta_B$ fixed to the WMAP estimate.

\begin{table}[!ht]
\renewcommand{\arraystretch}{1.3}
\caption{Theoretical abundances in the standard model} 
\label{standard-abundances}
\begin{tabular}{|c|c|}
\hline Nucleus & Our Code  \\ \hline
$\De$&$2.569 \times 10^{-5}$
\\  \hline 
$\He$&$0.248$\\ \hline
$\Li$&$4.514 \times 10^{-10}$\\ \hline
\end{tabular}
\end{table}

In order to check the consistency of the data, we follow the analysis
of \citet{PDGBook} for the data set considered in this work. We find
that the ideogram method plots are not Gaussian like, suggesting the
existence of unmodelled systematic errors. We take them into account
by increasing the standard deviation by a factor $S$. The values of
$S$ are $2.10$, $1.40$ and $1.90$ for $\De$, $\He$ and $\Li$
respectively. A scaling of erros was also suggested by \citet{olive07}.

We compute the light nuclei abundances and perform a statistical
analysis to obtain the best fit values for the parameters, for two
different cases: 
\begin{itemize}
\item variation of $\alpha$ allowing $\eta_B$ to vary,
\item variation of $\alpha$ keeping $\eta_B$ fixed.
\end{itemize}

Even though the WMAP data are able to constraint the baryon density
with great accuracy, there is still some degeneracy between the
parameters involved in the statistical analysis. For this reason, we
allow the joint variation of baryon density and the fine structure
constant to obtain an independent estimation for $\eta_B$.

\subsection{Variation of $\alpha$ and $\eta_B$}

In this case, we compute the light nuclei abundances for different values of
$\eta_B$ and $\frac{\Delta \alpha}{\alpha_0}$ and perform a
statistical analysis in order to obtain the best fit values for these
parameters. As pointed out by several authors
\citep{cuoco03,cyburt04,Vangioni04,ichi02,Ichi04}, there is no good
fit for the whole data set even for $\frac{\Delta \alpha}{\alpha_0}
\neq 0$. However, reasonable fits can be found excluding one group of
data at each time (see table \ref{resulbbn1}).

\begin{figure}[!ht]
\begin{center}
\includegraphics[scale=0.57,angle=0]{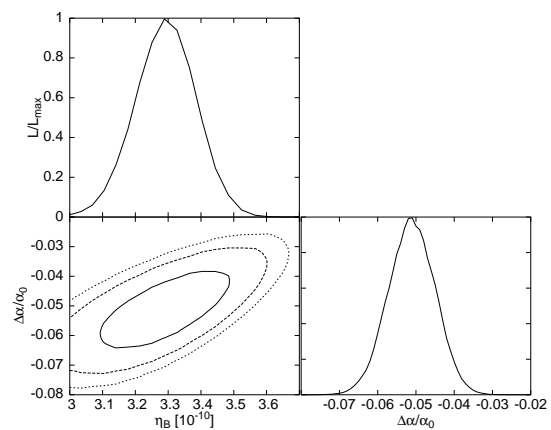}
\end{center}
\caption{$1\sigma$, $2\sigma$ and $3\sigma$ likelihood contours for
$\frac{\Delta \alpha}{\alpha_0}$ vs $\eta_B$ (in units of $10^{-10}$)
  and 1 dimensional Likelihood using the data of 
$\He + \Li$.}
\label{nucleo1}
\end{figure}

\begin{figure}[!ht]
\begin{center}
\includegraphics[scale=0.57,angle=0]{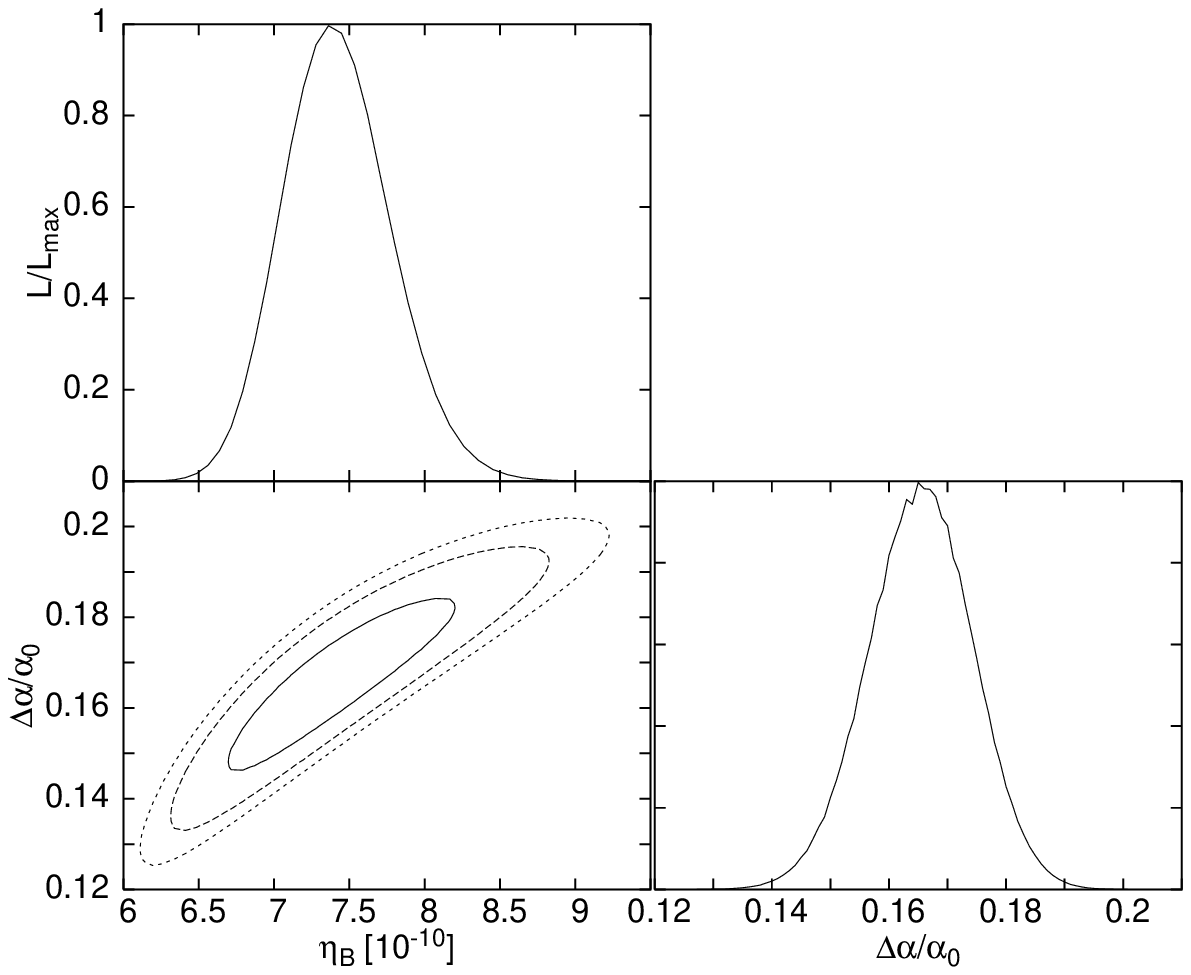}
\caption{$1\sigma$, $2\sigma$ and $3\sigma$ likelihood contours for
$\frac{\Delta \alpha}{\alpha_0}$ vs $\eta_B$ (in units of $10^{-10}$)
  and 1 dimensional Likelihood using the data of 
 $\De + \Li$.}
\label{nucleo2}
\end{center}
\end{figure}

\begin{figure}[!ht]
\begin{center}
\includegraphics[scale=0.57,angle=0]{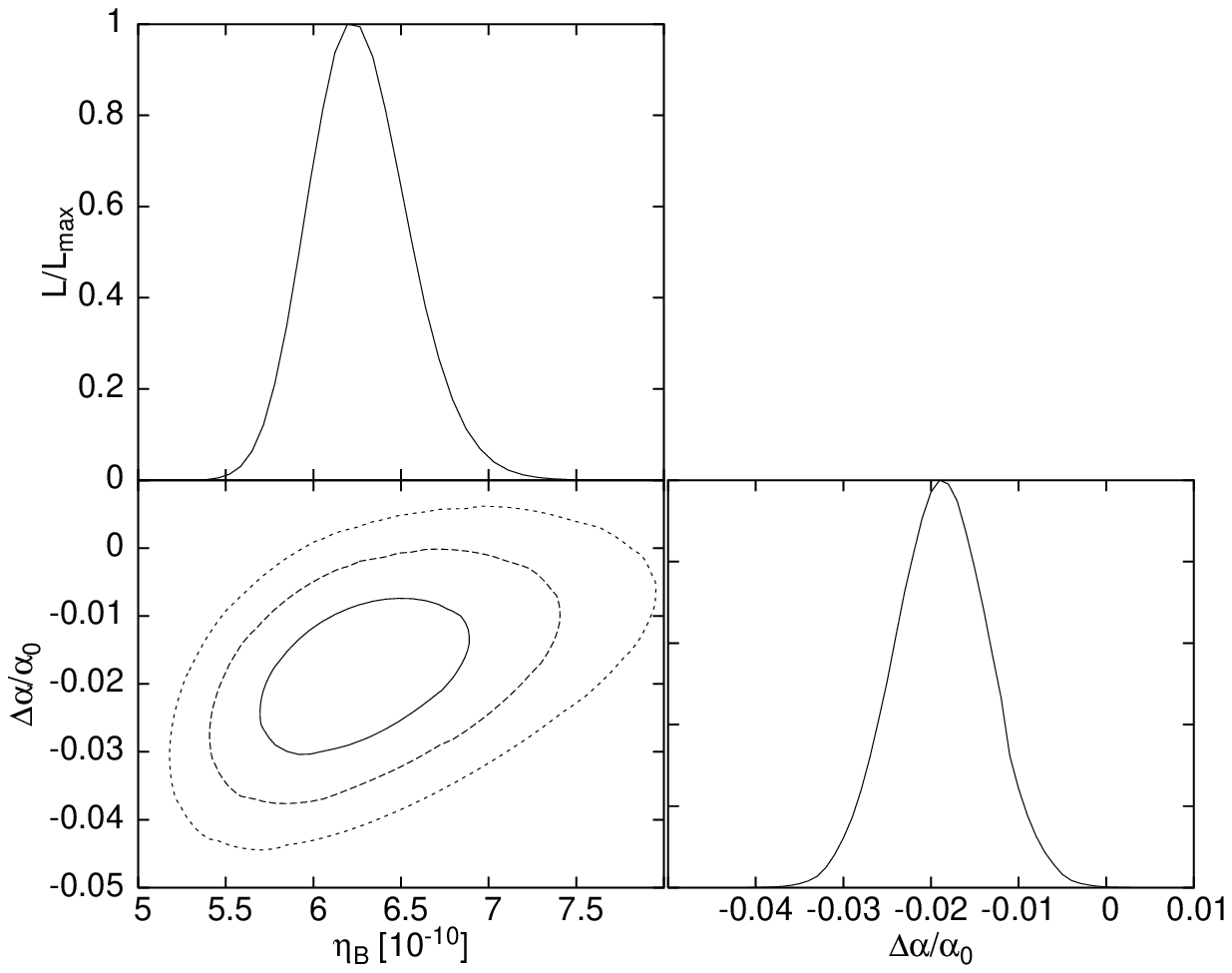}
\caption{$1 \sigma$, $2 \sigma$ and $3 \sigma$ likelihood contours for
$\frac{\Delta \alpha}{\alpha_0}$ vs $\eta_B$ (in units of $10^{-10}$)
  and 1 dimensional Likelihood using the data of  $\De + \He$.}
\label{nucleo3}
\end{center}
\end{figure}

For $\De + \He$, the value of $\eta_B$ is coincident with WMAP
estimation and there is no variation of $\alpha$ within $3\sigma$. On
the other hand, the other groups of data, favour values far from WMAP
estimation, and for $\De + \Li$, the result is consistent with
variation of $\alpha$ within $6\sigma$. In figures \ref{nucleo1} to 
\ref{nucleo3} the confidence contours and 1 dimensional
Likelihoods are shown, considering the available data of 
$\He +\Li$, $\De + \Li$ and $\De + \He$, respectively. 

\begin{table}[!ht]
\renewcommand{\arraystretch}{1.3}
\begin{center}
\caption{Best fit parameter values and $1\sigma$ errors for the BBN
constraints on $\frac{\Delta\alpha}{\alpha_0}$ and $\eta_B$ (in units of $10^{-10}$).}
\label{resulbbn1}
\begin{tabular}{|c|c|c|c|}
\hline  Data & $\eta_B \left[ 10^{-10}\right]$&$\frac{\Delta
\alpha}{\alpha_0}$&$\frac{\chi^2_{min}}{N-2}$ \\ \hline 
$\De + \He + \Li$& 
$4.188_{-0.095}^{+0.098}$& $-0.008_{-0.007}^{+0.004}$ & 10.33
\\ \hline 
$\He + \Li$ & 
$3.289_{-0.148}^{+0.135}$ & $-0.051 \pm 0.009$&  1.00
\\ \hline 
$\De + \Li$ & 
$7.362_{-0.491}^{+0.572}$ & $ 0.165 \pm 0.012$&  1.00
\\ \hline
$\De + \He$ & 
$6.195_{-0.418}^{+0.443}$& $-0.019 \pm 0.008$&  1.00
\\ \hline
\end{tabular}
\end{center}
\end{table}

\subsection{Variation of $\alpha$ with $\eta_B$ fixed}

Once again, we compute the light nuclei abundances for different values of
$\frac{\Delta \alpha}{\alpha_0}$, keeping $\eta_B$ fixed to the 
WMAP estimation \citep{wmap3}, and perform a statistical analysis in order to obtain the
best fit value for $\frac{\Delta \alpha}{\alpha_0}$. As pointed out in the previous
section, there is no good fit for the whole data set and for $\He +\Li$, even for
$\frac{\Delta \alpha}{\alpha_0} \neq 0$ (see table \ref{resulbbn2}).

\begin{figure}[!ht]
\begin{center}
\includegraphics[scale=0.3,angle=-90]{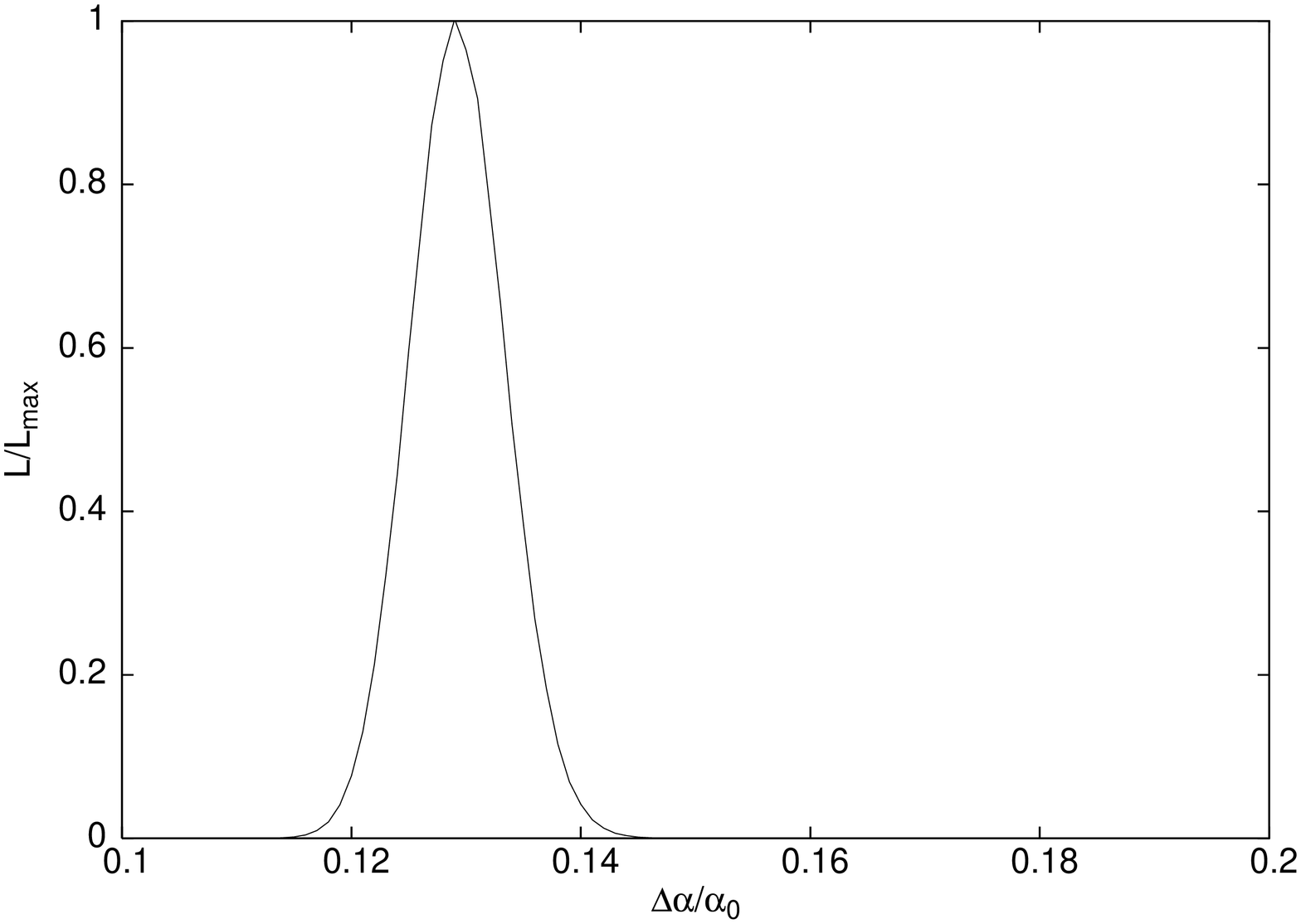}
\caption{1 dimensional Likelihood of $\frac{\Delta \alpha}{\alpha_0}$ using the data of
$\De + \Li$.}
\label{nucleo4}
\end{center}
\end{figure}

\begin{figure}[!ht]
\begin{center}
\includegraphics[scale=0.3,angle=-90]{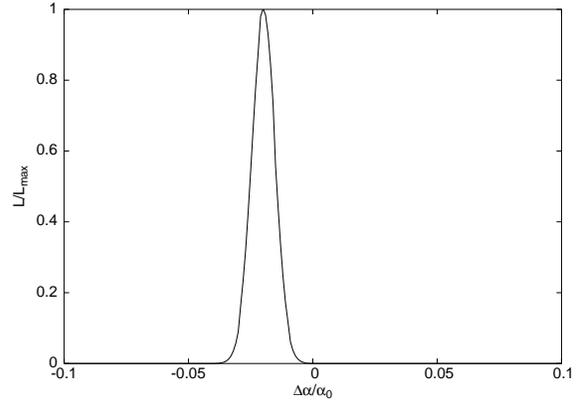}
\caption{1 dimensional Likelihood of $\frac{\Delta \alpha}{\alpha_0}$ using the data of
 $\De + \He$.}
\label{nucleo5}
\end{center}
\end{figure}

For $\De + \Li$, the result is consistent with variation of $\alpha$
within $6\sigma$, meanwhile for $\De + \He$ there is no variation of
$\alpha$ within $3\sigma$. In figures \ref{nucleo4} and  
\ref{nucleo5} the 1 dimensional Likelihood is shown, considering the
available data of $\De + \Li$ and $\De + \He$, respectively.  

\begin{table}[!ht]
\renewcommand{\arraystretch}{1.3}
\begin{center}
\caption{Best fit parameter values and $1\sigma$ errors for the BBN
constraints on $\frac{\Delta\alpha}{\alpha_0}$.}
\label{resulbbn2}
\begin{tabular}{|c|c|c|}
\hline  Data &$\frac{\Delta \alpha}{\alpha_0}$&$\frac{\chi^2_{min}}{N-1}$ \\ \hline 
$\De + \He + \Li$& $0.077 \pm 0.001$ & 23.28
\\ \hline 
$\He + \Li$ & $0.077 \pm 0.001$&  45.18
\\ \hline 
$\De + \Li$ & $ 0.129 \pm 0.006$&  1.58
\\ \hline
$\De + \He$ &  $-0.020 \pm 0.007$&  0.90 
\\ \hline
\end{tabular}
\end{center}
\end{table}

In order to test Bekenstein model, we consider the results
obtained using $\De + \He$ in this section.

It is worth to mention that if we don't consider $\Li$ data the
results with and without varying $\eta_B$ are the same. 

\section{Bounds from CMB}
\label{cmb}

Cosmic Microwave Background (CMB) radiation provides valuable
information about the physical conditions of the universe just before
decoupling of matter and radiation, and thanks to its dependence upon
cosmological parameters, it allows their estimation.  Any change in
the value of the fine structure constant affects the physics during
recombination, mainly the redshift of this epoch, due to a shift in
the energy levels and in particular, the binding energy of Hydrogen.
The Thompson scattering cross section is also changed for all
particles, being proportional to $\alpha^2$. Therefore, the CMB power
spectrum is modified by a change in the relative amplitudes of the
Doppler peaks, and shifts in their positions.  On the other hand,
changes in the cosmological parameters produce similar effects.
Previous analysis of the CMB data including a possible variation of
$\alpha$ have been performed by \citet{Martins02, Rocha03, Ichi06}. In
this paper, we use the WMAP 3-year temperature and
temperature-polarization power spectrum \citep{wmap3}, and other CMB
experiments such as CBI \citep{CBI04}, ACBAR \citep{ACBAR02}, and
BOOMERANG \citep{BOOM05_polar,BOOM05_temp}, and the power spectrum of
the 2dFGRS \citep{2dF05}. We consider a spatially-flat cosmological
model with adiabatic density fluctuations. The parameters of our model
are:
\begin{equation}
P=(\Omega_B h^2, \Omega_{CDM} h^2, \Theta, \tau, \frac{\Delta \alpha}{\alpha_0}, n_s, A_s)
\end{equation}
where $\Omega_{CDM} h^2$ is the dark matter density in units of the
critical density, $\Theta$ gives the ratio of the comoving sound
horizon at decoupling to the angular diameter distance to the surface
of last scattering, $\tau$ is the reionization optical depth, $n_s$
the scalar spectral index and $A_s$ is the amplitude of the density
fluctuations.

\begin{figure*}[!ht]
\begin{center}
\includegraphics[scale=0.7,angle=-90]{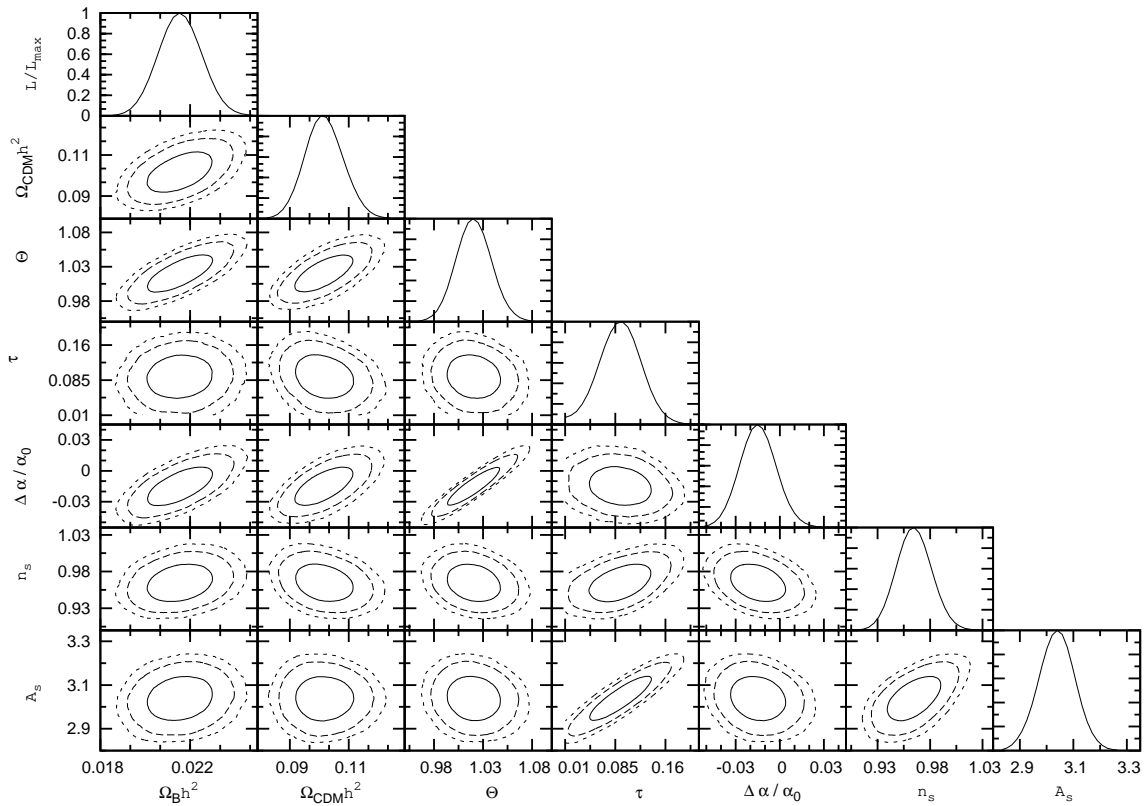}
\caption{Marginalized posterior distributions obtained with CMB data,
  including the WMAP 3-year data release plus 2dFGRS power
  spectrum. The diagonal shows the posterior distributions for
  individual parameters, the other panels shows the 2D contours for
  pairs of parameters, marginalizing over the others.}
\label{resulcmb}
\end{center}
\end{figure*}

We use a Markov Chain Monte Carlo method to explore the parameter
space because the exploration of a multidimensional parameter space
with a grid of points is computationally prohibitive.  We use the
public available CosmoMC code of \citet{LB02} which uses CAMB
\citep{LCL00} and RECFAST \citep{recfast} to compute the CMB power
spectra, and we have modified them in order to include the possible
variation of $\alpha$ at recombination. We ran eight different
chains. We used the convergence criterion of \citet{Raftery&Lewis} to
stop the chains when $R-1 < 0.003$ which is more stringent than the usually
adopted value. Results are shown in table \ref{tablacmb} and
figure \ref{resulcmb}.  Figure \ref{resulcmb} shows a strong
degeneracy between $\alpha$ and $\Theta$, which is directly related to
$H_0$, and also between $\alpha$ and $\Omega_B h^2$.  The values
obtained for $\Omega_B h^2$, $h$, $\Omega_{CDM} h^2$, $\tau$, and
$n_s$ are in agreement, within $1\sigma$, with the respective values
obtained without including any variation of $\alpha$ by the WMAP team
\citep{wmap3}.  Our results are consistent within $2 \sigma$ with no
variation of $\alpha$ at recombination.

\begin{table}[!ht]
\renewcommand{\arraystretch}{1.3}
\begin{center}
\caption{Mean values and errors for the principal and derived parameters including $\alpha$ variation.}
\label{tablacmb}
\begin{tabular}{|c|c|}
\cline{1-2} 
  Parameter & Mean value and $1\sigma$ error \\
\cline{1-2} 
$\Omega_B h^2$ & 0.0215$_{-0.0009}^{+0.0009}$ \\
\cline{1-2}
$\Omega_{CDM} h^2$ & 0.102$_{-0.006}^{+0.006}$ 
\\
\cline{1-2}
$\Theta$ & 1.021$_{-0.017}^{+0.017}$ \\
\cline{1-2}
$\tau$ & 0.092$_{-0.014}^{+0.014}$ 
\\
\cline{1-2} 
$\Delta \alpha / \alpha_0$ & -0.015$_{-0.012}^{+0.012}$ \\
\cline{1-2}
$n_s$ & 0.965$_{-0.016}^{+0.016}$ 
\\
\cline{1-2} 
$A_s$ & 3.039$_{-0.065}^{+0.064}$ \\
\cline{1-2}
$H_0$ &  67.7$_{-4.6}^{+4.7}$ 
\\
\cline{1-2}
\end{tabular}
\end{center}
\end{table}

\begin{figure}[!ht]
\begin{center}
\includegraphics[scale=0.5,angle=-90]{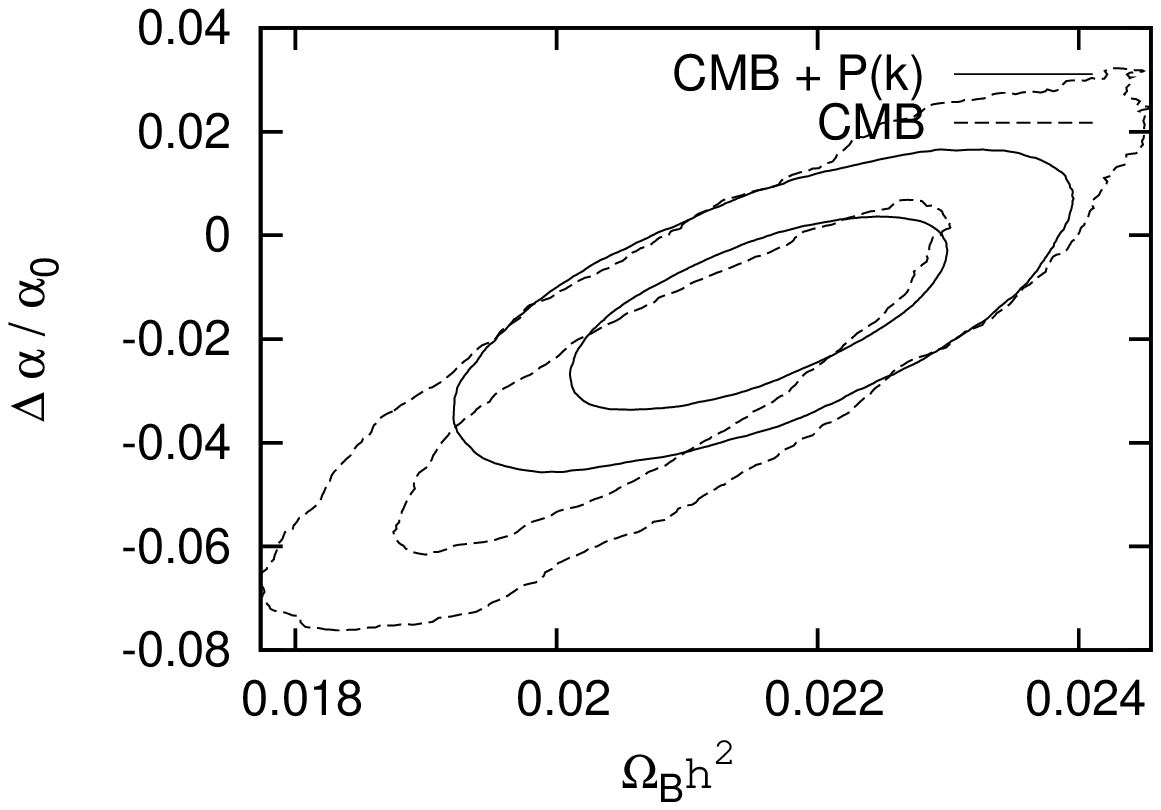}
\end{center}
\caption{$1\sigma$ and $2 \sigma$ confidence levels contours obtained
  with CMB data with and without data of the 2dFGRS power spectrum.}
\label{individuales}
\end{figure}

\begin{figure}[!ht]
\begin{center}
\includegraphics[scale=0.5,angle=-90]{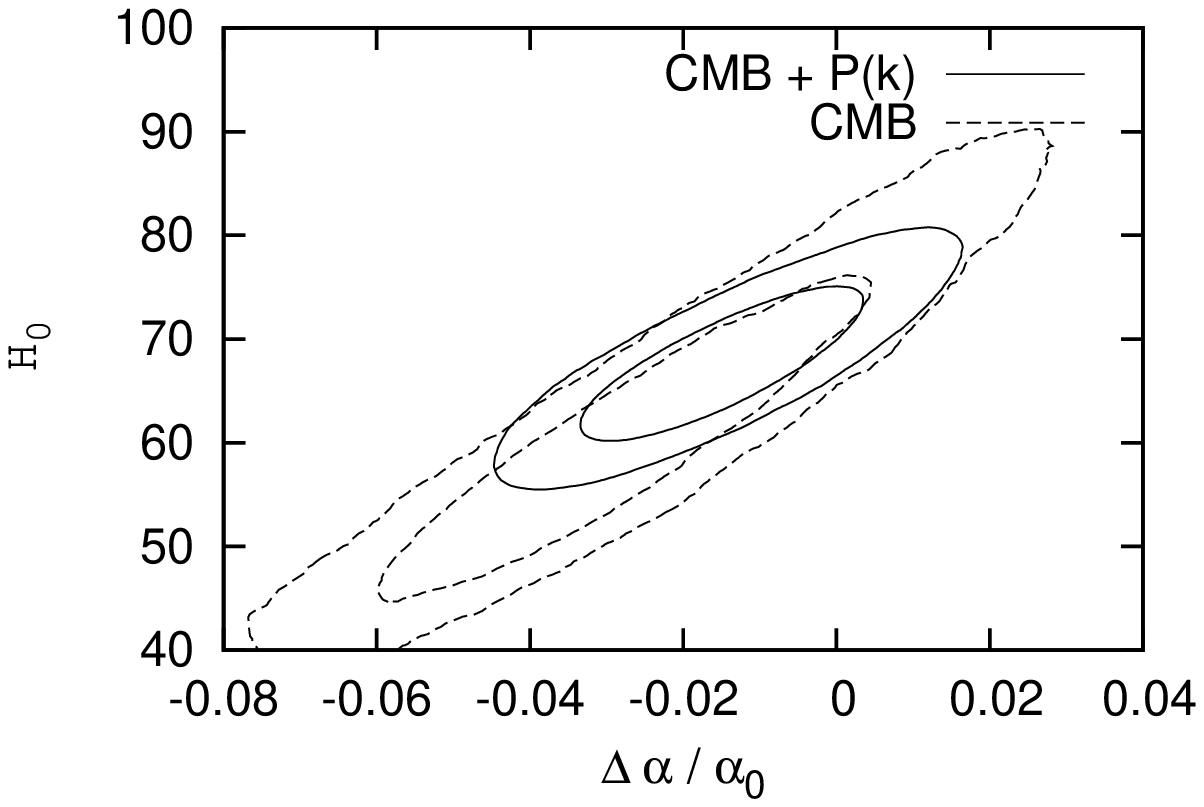}
\end{center}
\caption{$1\sigma$ and $2 \sigma$ confidence levels contours obtained
  with CMB data with and without data of the 2dFGRS power spectrum.}
\label{individuales2}
\end{figure}

We have also performed the analysis considering only CMB data. In that
case, the strong degeneracy between $\alpha$ and $H_0$ made the chains
cover all the wide $H_0$ prior, making it impossible to find reliable
mean values and errors. Hence, we added a Gaussian prior to $H_0$, which
was obtained from the Hubble Space Telescope Key Project
\citep{hst01}, and chose the values of the mean and errors as those
inferred from the closest objects in that paper, so we could neglect
any possible difference between the value of $\alpha$ at that redshift
and the present value. In this way, we post-processed the chains, and
found that the most stringent constraints were obtained in the first
analysis (see figures \ref{individuales} and \ref{individuales2}).

\section{Bounds from Quasar Absorption Systems}
\label{quasars}

Quasar absorption systems present ideal laboratories to search for any
temporal variation in the fine structure constant. Quasar spectra of
high redshift show the absorption resonance lines of the alkaline ions
like CIV, MgII, FeII, SiIV and others.  The relative magnitude of the
fine splitting of resonance lines of alkaline ions is proportional to
$\alpha^2 $. Several authors
\citep{CS95,VPI96,Murphy01b,chand05,MVB04} studied the SiIV doublet
absorption lines systems at different redshifts ($2.5 < z < 3.33$), to
put bounds on the variation of $\alpha$.  \citet{Bahcall04} used O III
emission lines.  \citet{Webb99,Murphy01a,Murphy03b} compared
transitions of different species with far between atomic masses
and led to a single data consistent with time varying fine structure
constant for a range of redshifts ($0.5 < z < 3.5$).  However, other
recent independent analysis of similar data
\citep{QRL04,Srianand04,grupe05} found no variation. Another method,
to test cosmological variation of $\alpha$, was proposed by
\citet{levshakov05} from pairs of Fe II lines observed in individual
exposures from a high resolution spectrograph. The authors found no
variation of $\alpha$ at $z=1.15$ and $z=1.839$. However, a recent
reanalysis of spectrum of the quasar Q1101-264 found variability
within $1\sigma$ \citep{Levshakov07}.  We also consider in our analysis
the bounds mentioned in \citet{WBR76,SM79,CS95,tzana07}, which were
obtained by comparing the optical and radio redshifts.  Furthermore,
\citet{Murphy02} compare molecular and radio lines and obtain a more
stringent constraints. On the other hand, \citet{darling04} reports
bounds on the variation of $\alpha$ at $z=0.2467$ from the satellite
$18$ cm $OH$ conjugate lines. Finally, \citet{kanekar05} compared the
HI and OH main line absorption redshifts of the different components
in the $z=0.765$ absorber and the $z=0.685$ lens toward B0218+357 to
place stringent constraints on changes in $F= g_p
\left({\frac{\alpha^2}{\mu}}\right)^{1.57}$.

Since we want to compare the prediction of $\alpha$ evolution with
cosmological time, we consider each individual measurement of the
papers cited above and not the average value reported in each paper.

\section{Bounds from Geophysical Data}
\label{geo}

\subsection{The Oklo Phenomenon}

One of the most stringent limits on time variation of the fine
structure constant follows from the analysis of isotope ratios
in the natural uranium fission reactor that operated $1.8\times 10^9$
years ago at the present day site of the Oklo mine in Gabon,
Africa. The proof of the past existence of a spontaneous chain
reaction in the Oklo ore consists essentially of a substantial
depletion of the uranium isotopic ratio $^{235}{\rm U}/^{238}{\rm U}$
with respect to the current standard value in terrestrial samples and a
correlated peculiar distribution of some rare-earth isotopes.  
From an analysis of nuclear and geochemical data, the
operating conditions of the reactor could be reconstructed and the
thermal neutron capture cross sections of several nuclear species be
measured. 

The large values of the thermal capture cross
sections of $^{149}{\rm Sm}$,$^{155}{\rm Gd}$ and $^{157}{\rm Gd}$ are
due to the existence of resonances in the thermal region. In presence
of such resonance, the mono-energetic capture cross section is well
described in the thermal region by the  Breit-Wigner formula: 
\begin{equation}
\sigma_{n,\gamma}=\pi \frac{\hbar ^2}{p^2} g
\frac{\Gamma_n\left(E\right) \Gamma_{\gamma}}{{\left(E - E_r\right)}^2
  + \frac{\Gamma^2}{4}} \nonumber 
\end{equation}
where $p$ is the momentum of the neutron, $g$ a statistical factor
depending upon the spins of the compound nucleus of the incident
neutron and target nucleus, $\Gamma_{\gamma}$ is the radiative
width, ${\Gamma}_n\left(E\right)$ is the neutron partial width,
${\Gamma}$ is the total width. Thus, a shift in the lowest lying resonance level 
in $^{149}{\rm Sm}: \Delta = E_r^{149{\rm(Oklo)}} -E_r^{149{\rm(now)}}$ 
can be derived from a shift in the neutron capture cross section of
the same nucleus \citep{Fujii00,DD96}. The shift in the resonance
energy can be translated \citep{DD96} into bounds on a 
possible difference between the value of $\alpha $ during the Oklo
phenomenon and its actual value. 

Various authors \citep{DD96,Fujii00,LT04} have analyzed the Oklo data in order to put
bounds on $\alpha$. \citet{Fujii00} used samples of
$^{149}{\rm Sm}$, $^{155}{\rm Gd}$ and $^{157}{\rm Gd}$ to reanalyze
the bound on the resonance energy. They took the effect of
contamination into account, assuming the same contamination parameter
for all samples. \citet{LT04} employ a more realistic spectrum than
the commonly used Maxwell-Boltzmann to put the following bound on
$\alpha$: 
\begin{equation}
\frac{\Delta \alpha}{\alpha_0} = (-45 \pm 15) \times 10^{-9} 
\end{equation}

This bound is very similar to the one found by \citet{Fujii00} and
therefore we are going to consider it when testing the Bekenstein
model. 

\subsection{Long-Lived $\beta$ Decayers}

The half-life of long-lived $\beta $ decayers such $^{187}\rm{Re}$ has
been used by several authors to find bounds on the variation of
$\alpha$. These nuclei have a very long half-life that has been
determined either in laboratory measurements or by comparison with the
age of meteorites. This last quantity can be measured from
$\alpha $ decay radioactivity analysis. The most stringent bound on
the variation of the half life, $\lambda$, proceeds from the
comparison of $^{187}\rm{Re}$ decay in the Solar System formation and
the present \citep{Olive04b}: 
\begin{equation}
\frac{\Delta \lambda}{\lambda} = (-0.016 \pm 0.016) 
\end{equation}
\citet{SV90} derived a relation between the
shift in the half-life of long lived $\beta $ decayers and a possible
variation between the values of the fundamental constants $\alpha,
\Lambda _{QCD}$ and $G_F$ at the age of the meteorites and their
values now. They use a phenomenological model in which the abundance
of any unstable nucleus will obey the following decaying law: 
\begin{equation}
N = N_0 \exp\left[-\left(\lambda t+ \dot \lambda t^2/2\right)\right]
\end{equation} 
In this paper, we only consider $\alpha$ variation and
therefore, the following equation holds:
\begin{equation}
\frac{\Delta \lambda}{\lambda} = a \frac{\Delta \alpha}{\alpha_0}
\end{equation}
where $a=21600$ for $^{187}\rm{Re}$.

\section{Bounds from Laboratory}
\label{lab}
The comparison of different atomic transition frequencies over time
can be used to determine the present value of the temporal derivative
of $\alpha$. Indeed, the most stringent limits on the variation of
$\alpha$ are obtained using this method. The dependence of hyperfine
transition frequencies with $\alpha$ can be expressed as: 
\begin{equation} 
\nu_{Hyp} \sim \alpha^2 \frac{\mu}{\mu_B}\frac{m_e}{m_p} R_{\infty} c
F_{REL}(\alpha Z) 
\end{equation}
where $\mu$ is the nuclear magnetic moment, $\mu_B$ is Bohr magneton,
$R_{\infty}$ is Rydberg's constant, $m_p$ and $m_e$ are the proton and
electron mass and $F_{REL}$ is the relativistic contribution to the
energy. 

The comparison of rates between clocks based on
hyperfine transitions in alkali atoms with different atomic number $Z$
can be used to set bounds on $\alpha^k \frac{\mu_{A_1}}{\mu_{A_2}}$
where $k$ depends on the frequencies measured and $\mu_{A_i}$ refers
to the nuclear magnetic moment of each atom. The first three entries
of table \ref{clocks} show the bounds on
$\frac{\dot{\alpha}}{\alpha_0}$ obtained by comparing hyperfine
transition frequencies in alkali atoms. 

On the other hand, an optical transition frequency
has a different dependence on $\alpha$:
\begin{equation}   
\nu_{opt} \sim R_{\infty} B F_i(\alpha)
\end{equation}
where $B$ is a numerical constant assumed not to vary in time and
$F_i(\alpha)$ is a dimensionless function of $\alpha$ that takes into
account level shifts due to relativistic effects. The comparinson between an
optical transition frequency and an hyperfine transition frequency
can be used to set bound on $\alpha^k \frac{m_e}{m_p}
\frac{\mu_A}{\mu_B}$. 

Different authors, \cite{Bize03,Fischer04,Peik04}, have
measured different optical transitions and set bounds on the variation
of $\alpha$ using different methods. 
\citet{Fischer04} have considered the joint variation of $\alpha$ and
$\frac{\mu_{Cs}}{\mu_B}$. We have reanalyzed the data of
\citet{Fischer04}, considering only $\alpha$ variation, yielding
the fifth entree of table \ref{clocks}.  On the other hand, 
\citet{Peik04} have measured an optical transition frequency in
$^{171}{\rm Yb} $ with a cesium atomic clock. They perform a linear regression
analysis using this result together with other optical transition
frequency measurements from \citet{Bize03,Fischer04}. 
We have already considered the other data, therefore, we have
reanalyzed the data, using only the comparison between Yb and Cs
frequency, yielding the sixth entree in table \ref{clocks}.

\begin{table*}[!ht]
\renewcommand{\arraystretch}{1.3}
\begin{center}
\caption{The table shows the compared clocks, the value of
  $\frac{\dot \alpha}{\alpha_0}$ and its corresponding error in units
  of $10^{-15} {\rm yr}^{-1}$, the
time interval for which the variation was measured and the reference. References (1) \citet{PTM95}; (2) \citet{Sortais00}; (3) \citet{Marion03}; (4) \citet{Bize03}; (5) \citet{Fischer04}; (6) \citet{Peik04}}
\begin{tabular}{|c|c|c|c|}
\hline
 Frequencies  & $\frac{\dot \alpha}{\alpha_0} \pm \sigma
\left[ 10^{-15} \rm{yr}^{-1}\right] $  & $\Delta t [\rm{yr}]$& 
Reference \\ \hline
Hg+ and   H maser & $0.0 \pm 37.0$ & 0.38  &  (1) \\\hline
 Cs and Rb & $4.2 \pm 6.9 $& 2 & (2)\\\hline
 Cs and Rb & $-0.04 \pm 1.60 $ & 5 & (3) \\\hline
Hg and Cs & $ 0.0 \pm 1.2 $ & 2 & (4) \\\hline
H and Cs & $1.14 \pm 2.25 $ & 5 & (5) \\\hline
Yb and  Cs & $-0.58  \pm 2.1 $ & 2.8 & (6) \\\hline
\end{tabular}
\end{center}
\label{clocks}
\end{table*}

\section{The Model}
\label{modelo}

In this section, we solve the equation of the scalar field, which
drives the variation of $\alpha$ in the Bekenstein model. First, we
obtain the analytical solution for Friedmann-Robertson-Walker (FRW)
equation for two different regimes and assure continuity of the solution and its
derivative. Unlike other works \citep{BSM02,OP02}, we do not assume
that the scalar field is connected with dark matter field. 
We consider the weak field approximation and so only the
electrostatic contribution to the scalar field equation is
relevant. In this framework, the electric charge can be expressed in
the form: 
\begin{equation}
e=e_0 \epsilon(x^\mu)
\end{equation}
being $\epsilon$ a scalar field. The Lagrangian for a charged particle
of rest mass $m$ and charge $e_0 \epsilon$:
\begin{equation}
L=-m c \left(u^\mu u_\mu\right)^{1/2} + \frac{e_0
\epsilon}{c} u^\mu A_\mu \label{lag}
\end{equation}
where $u^\mu=\frac{d x^\mu}{d\tau}$. $L$ is Lorentz invariant and
after a gauge transformation it changes only 
by a perfect derivative. Following \citet{Bekenstein82}, we obtain the Lagrange equations for (\ref{lag}):
\begin{equation}
\frac{{\rm d}\left(m u_\nu\right)}{{\rm d} \tau}=\frac{e_0}{c}
\left\{\left(\epsilon A_\mu\right)_{,\,\nu}-\left(\epsilon
A_\nu\right)_{,\, \mu}\right\}u^\mu -m_{,\, \nu} c^2
\end{equation}
And identify $F_{\mu\nu}$: 
\begin{equation}
F_{\mu \nu} = \epsilon^{-1} \left[\left(\epsilon A_{\nu}\right),_{\mu}
  -\left(\epsilon A_{\mu}\right),_{\nu} \right] 
\end{equation}

Following \citet{Bekenstein82}, the total action can be written
as: 
\begin{equation}
S=S_m+S_{EM}+S_\varepsilon + S_g
\end{equation}
where 
\begin{equation}
S_g = \frac{c^4}{16 \pi G} \int \sqrt{(-g)} R d^4x
\end{equation}
belongs to the gravitational sector of the theory, 
\begin{equation}
S_{EM}= -{\left(16 \pi\right)}^{-1} \int F^{\mu \nu} F_{\mu \nu} {(-g)}^{1/2} d^4x
\end{equation}
is the electromagnetic action and 
\begin{equation}
S_m = \sum_i \int \frac{1}{\gamma}\left[-m c^2 +(e_0 \epsilon) u^{\mu} A_{\mu}\right]
 \delta^3\left[x^i - x^i (\tau)\right] d^4x 
\end{equation}
is the matter action where the coupling between matter and the gauge
field depends on the scalar field responsible for the variation of
$\alpha$. The action of the scalar field can be expressed as: 
\begin{equation}
S_\epsilon = -\frac{1}{2} \frac{\hbar c}{l^2} \int
\frac{\epsilon_{,\mu}\epsilon^{,\mu}}{\epsilon^2} {(-g)}^{1/2} d^4x 
\end{equation}
where $l$ is a scale length which is introduced due to dimensional
reasons and one of the assumption of this theoretical framework is $l
> L_p$.  It will be shown (see apendix \ref{eotvos}) that this  
latter condition implies violation of the weak equivalence principle.  However, this requirement could be relaxed, due to string theories
considerations \citep{strings1,strings2}. 

Varying the total action with respect to the gauge field,
the modified Maxwell equations are obtained:
\begin{equation}
\left(\epsilon^{-1} F^{\mu \nu}\right)_{;\, {\nu}} =  4 \pi j^{\mu}
\end{equation}
where
\begin{equation}
j^{\mu} = \sum_i \frac{e_0}{c \gamma} u^{\mu} \frac{1}{\sqrt {-g}}
\delta^3\left[x^i - x^i\left(\tau\right)\right] 
\end{equation}
as well as the equation of motion of the scalar field:
\begin{equation}
\Box \ln \epsilon = \frac{l^2}{\hbar c} \left[ \epsilon \frac{\partial
    \sigma}{\partial \epsilon} - \frac{1}{8 \pi} F^{\mu \nu} F_{\mu
    \nu} \right]  \label{e1}
\end{equation}
where
\begin{equation}
\sigma = \sum _ i \frac{m c^2}{\gamma} \frac{1}{\sqrt {-g}
}\delta^3\left[x^i - x^i\left(\tau\right)\right] 
\end{equation}
In an expanding Universe and evaluating the r.h.s of equation (\ref{e1}) following
\citet{Bekenstein82}, the following expression can be obtained: 
\begin{eqnarray}
\label{de}
\partial_t \left(\frac{\partial_t{\epsilon}}{\epsilon} a^3(t)\right)&=& -a^3(t)
\frac{l^2}{\hbar c}\zeta \rho_m c^4
\end{eqnarray}
where $\zeta=\frac{\rho_{em}}{\rho_m}$ is a dimensionless parameter
which measures the fraction of mass in the form of Coulomb energy to
the total matter density $\rho_m$. As suggested in \citet{SBM02} we
use $\zeta=10^{-4}$. Integrating equation (\ref{de}), in an expanding
universe and using $\rho_m= \frac{ \Omega_m \rho_c}{a^3(t)}$, we
obtain: 
\begin{eqnarray}
\label{importante} \frac{\partial_t{\epsilon}}{\epsilon} &=&-\frac{3}{8
\pi} \zeta \left(\frac{l}{L_p}\right)^2 H_0^2 \Omega_m
\left(\frac{a_0}{a(t)}\right)^3 (t-t_c)
\end{eqnarray}
where $t_c$ is an integration constant and $\Omega_m$ is the total
matter density in units of the critical density ($\rho_c$). In order to
solve the above equation we must first solve the Friedmann
equation for the different regimes we are considering.

In a flat Friedman-Robertson-Walker (FRW) universe, the equation for the scale factor reads:
\begin{eqnarray}
\label{friedmann} \left(\frac{\partial_t{a}}{a}\right)^2 &=& H_0^2
\left\{\Omega_m \left(\frac{a_0}{a(t)}\right)^3+\Omega_r
\left(\frac{a_0}{a(t)}\right)^4+\Omega_\Lambda \right\}
\end{eqnarray}
with the initial condition $ a\left( 0\right)=0$, and satisfying that
$a\left(t_0\right)=a_0=1$. In the above equation, we assume that the
scalar field contribution is negligible. Usually, this contribution is
proportional to $\left(\frac{\partial_t \epsilon}{\epsilon}\right)^2$ and we
expect the variation of $\alpha$ to be of order $10^{-5}$. 

The FRW equation has no analytical solution in terms of elementary functions
when radiation, matter and cosmological constant are considered. We
build a piecewise approximate solution by joining solutions obtained
by conserving only some terms of the r.h.s of equation
(\ref{friedmann}). We solve the FRW equation for two different cases: a)
radiation and matter and b) matter and cosmological constant. In such
way, solution a) can be applied to nucleosynthesis and recombination
of primordial hydrogen whereas solution b) is proper for quasar
absorption systems, geophysical data and atomic clocks.

First, we integrate equation (\ref{friedmann}) considering only matter
and radiation. In order to get an analytical expression for the scale
factor as a function of time, we change the independent variable to
conformal time $\eta$ as follows: $a_{RM} {\rm d} \eta = {\rm
d}t$. Defining $\xi = H_0 \eta$, we can write:
\begin{eqnarray}
\label{adeambos}a_{RM}(\xi) &=&\frac{\xi^2 \Omega_m}{4}  + \xi
\sqrt{\Omega_r}
\end{eqnarray}
The time  can be expressed as follows: 
\begin{eqnarray}
\label{tdeambos} 
H_o t(\xi)&=&\frac{\xi^3 \Omega_m}{12}
 +\frac{\xi^2 \sqrt{\Omega_r}}{2}
\end{eqnarray}

Now, we solve equation (\ref{friedmann}) considering only matter and
cosmological constant and obtain:
\begin{eqnarray}
\label{amyc} a_{MC}(t) &=& \left(
\frac{\Omega_m}{\Omega_\Lambda}\right)^{1/3} \left[ {\rm sinh}
\left( \frac{3}{2} \sqrt{\Omega_\Lambda} H_0 \left(t-t_0 \right)\right. \right.  \nonumber \\ 
&& \hskip 2.5cm \left. \left. +{\rm arcsh} \sqrt{\frac{\Omega_\Lambda}{\Omega_m}}\right)\right]^{2/3}
\end{eqnarray}

The expansion factor must be a continuous and smooth
function of time, and in order to match both solutions, the following
conditions have to be fulfilled:
\begin{eqnarray}
a_{RM}(t_1)&=&a_{MC}(t_1) \nonumber \\
\frac{d a_{RM}}{d t}(t_1)&=&\frac{d a_{MC}}{d t}(t_1) \nonumber
\end{eqnarray}
from where, we obtain:
\begin{eqnarray}
a_{RM}(t_1)=a_{MC}(t_1)&=&\left(\frac{\Omega_r}{\Omega_\Lambda}\right)^{1/4}
\end{eqnarray}

In order to compare with astronomical and local bounds, we use the
value of the cosmological parameters reported in \citet{PDGBook}. 

Now we can solve equation (\ref{importante}) using the equations
(\ref{adeambos}) and (\ref{amyc}). Using $\ln \frac{\epsilon (t)}{\epsilon
(t_0)} \simeq \frac{1}{2} \frac{\Delta \alpha}{\alpha_0}$, we obtain
the following expressions for the variation of $\alpha$ in the two
different regimes.

Defining $\lambda(\xi)=\xi \Omega_m \ +4 \sqrt{\Omega_r}$ for $t<t_1$:
\begin{eqnarray}
\frac{\Delta \alpha}{\alpha_0}&=&-\frac{1}{\pi} \zeta \left(\frac{l}{L_p}\right)^2 {\rm ln}
\left(\frac{\lambda(\xi)}{\lambda(\xi_1)}\right)-\frac{1}{8\pi} \zeta
\left(\frac{l}{L_p}\right)^2 {\rm ln}
\left(\frac{\Omega_r}{\Omega_\Lambda} \right) \nonumber \\ 
&&+\frac{2}{\pi} \zeta \left(\frac{l}{L_p}\right)^2 \sqrt{\Omega_r} \left[\frac{1}{\lambda(\xi )}
-\frac{1}{\lambda(\xi_1 )}\right]\nonumber \\
&&-\frac{3}{4 \pi} \zeta \left(\frac{l}{L_p}\right)^2 \frac{\Omega_m
}{\Omega_r} H_0 t_c
\left\{\frac{1}{\xi}-\frac{1}{\xi_1}+\frac{\Omega_m}{\lambda(\xi
  )}\right. \nonumber \\ 
&&  \hskip 2cm \left. -\frac{\Omega_m}{\lambda(\xi_1 )}
+\frac{ \Omega_m}{2\sqrt{\Omega_r}}{\rm ln}\left[\frac{\xi \lambda(\xi_1 )}
{\xi_1 \lambda(\xi )}\right]\right\}\nonumber\\
&&+\frac{1}{2\pi} \zeta \left(\frac{l}{L_p}\right)^2
\sqrt{\Omega_\Lambda}\left\{\frac{H_0\left(t_1-t_c\right)}{{\rm
    th}\left({\rm arcsh}
  \sqrt{\frac{\Omega_\Lambda}{\Omega_m}\left(\frac{\Omega_r} 
{\Omega_\Lambda}\right)^{3/4}}\right)} \right. \nonumber \\  
\label{dalpharadiacion}
&& \hskip 3.3cm \left.-\frac{H_0\left(t_0-t_c\right)}{{\rm
    th}\left({\rm arcsh} \sqrt{\frac{\Omega_\Lambda}{\Omega_m}}
  \right)} \right\}  
\end{eqnarray}
Defining $\nu=H_0 t$, we can write for $t>t_1$:
\begin{eqnarray}
\frac{\Delta \alpha}{\alpha_0}&=&\frac{1}{2\pi} \zeta
\left(\frac{l}{L_p}\right)^2  \sqrt{\Omega_\Lambda} \times \nonumber\\ 
&&\left\{\frac{ \nu - \nu_c }{{\rm th}\left(\frac{3}{2}
  \sqrt{\Omega_\Lambda} \left(\nu-\nu_0 \right)+ {\rm
    arcsh}\sqrt{\frac{\Omega_\Lambda}{\Omega_m}}
  \right)}\right. \nonumber \\ 
&&\left.-\frac{\left(\nu_0-\nu_c\right)}{{\rm th}\left({\rm arcsh}
  \sqrt{\frac{\Omega_\Lambda}{\Omega_m}} \right)} \right\} \nonumber\\ 
&&-\frac{1}{3 \pi} \zeta \left(\frac{l}{L_p}\right)^2 {\rm ln}
\left[\sqrt{\frac{\Omega_m}{\Omega_\Lambda}}{\rm
    sinh}\left(\frac{3}{2} \sqrt{\Omega_\Lambda}  \left(\nu-\nu_0
  \right) \right. \right. \nonumber \\
\label{dalphamateria}
&& \hskip 2.7cm \left. \left. + {\rm arcsh}\sqrt{\frac{\Omega_\Lambda}{\Omega_m}} \right)\right] 
\end{eqnarray}

\section{Results}
\label{resultados}

\subsection{The early universe}

In section \ref{nucleo}, we have used the primordial abundances of
$\De$, $\He$ and $\Li$ to put bounds on the variation of $\alpha$ in
the early universe. First, we have performed a statistical analysis in
order to check the consistency of each group of data and modified the
observational errors accordingly. We have shown that all data could not
be fitted at the same time, but reasonable fits can be found
considering two groups of data at the time. We have analyzed the case
where the baryon density is a free parameter and the case where it is
fixed to the WMAP value. Tables and confidence contours are shown in
section \ref{nucleo}. In all analysis describe in section
\ref{nucleo} (with or without allowing $\eta_B$ to vary), we find that
excluding the $\Li$ data, our results are consistent with WMAP
estimation and no variation of $\alpha$.

In section \ref{cmb} we have used the three year WMAP data together
with other CMB experiments and the 2dFGRS power spectrum to put
constraints on the variation of $\alpha$ during recombination. Tables
and confidence contours are shown in section \ref{cmb}. 

We summarize our results of the variation of the fine structure constant
in the table \ref{resultados-sum}. Our results are consistent with no
variation of the fine structure constant in the early Universe.

\begin{table}[!ht]
\renewcommand{\arraystretch}{1.3}
\begin{center}
\caption{Best fit parameter values and $1\sigma$ errors for the BBN
  and CMB constraints on $\frac{\Delta\alpha}{\alpha_0}$.}
\label{resultados-sum}
\begin{tabular}{|c|c|}
\hline Group of Data & $\frac{\Delta \alpha}{\alpha_0}$ \\ \hline 
BBN& $-0.020 \pm 0.007$\\ \hline 
CMB& $-0.015 \pm 0.012$ \\ \hline
\end{tabular}
\end{center}
\end{table}

\subsection{The Bekenstein Model}

In this subsection we compare the Bekenstein model predictions obtained
in section \ref{modelo} with astronomical and geophysical data
described in sections \ref{quasars}, \ref{geo} and \ref{lab} and with the bounds
on $\alpha$ from the early universe we have obtained in sections
\ref{nucleo} and \ref{cmb}. 

Fixing the time, equation (\ref{dalpharadiacion}) or
(\ref{dalphamateria}) gives the prediction for $\alpha$ variation as a
function of two free parameters: $\left(\frac{l}{L_p}\right)^2$ and
$\left(\frac{l}{L_P}\right)^2 H_0 t_c $. Therefore, we have $N$
(number of data we are considering: 1 from Oklo, 1 from
$^{187}\rm{Re}$, 6 from atomic clocks, 1 from BBN, 1 from 
CMB, 274 from QSO) conditional equations with two unknowns. We perform
a $\chi^2$ test to obtain the best values of the free parameters of
Bekenstein's theory. Our results are shown in table \ref{table-beke}.
\begin{table}[!ht]
\renewcommand{\arraystretch}{2.}
\begin{center}
\caption{Best fit parameter values and $1\sigma$ errors of the
 Bekenstein model.} 
\label{table-beke}
\begin{tabular}{|c|c|c|}
\hline Data & $\left(\frac{l}{L_p}\right)^2$&$\left(\frac{l}{L_P}\right)^2
H_0 t_c$ \\ \hline
All data &$0.000\pm 0.003 $&$\left(3.2 \pm 1.4\right) \times 10^{-6}$\\ \hline
Without Oklo &$0.000\pm 0.014 $&$\left(3.2 \pm 1.4\right) \times 10^{-6}$\\ \hline
Without $^{187}\rm{Re}$&$0.000\pm 0.003 $&$\left(3.2 \pm 1.4\right) \times 10^{-6}$\\ \hline
Without atomic clocks&$0.000\pm 0.003 $&$\left(3.2 \pm 1.4\right) \times 10^{-6}$\\ \hline
Without BBN&$0.000\pm 0.017 $&$\left(3.4 \pm 1.3\right) \times 10^{-2}$\\ \hline
Without CMB&$0.000\pm 0.003 $&$\left(3.2 \pm 1.4\right) \times 10^{-6}$\\ \hline
\end{tabular}
\end{center}
\end{table}

We also have performed the same statistical analysis discarding bounds
of each group of data. In most of the cases, the results are similar
than those considering all data. However, discarding the bound from
nucleosynthesis changes the value of $\left(\frac{l}{L_P}\right)^2 H_0
t_c$ several orders of magnitude. Thus, the bound obtained from the
primordial abundances of the light elements are crucial to fix the
value of $\left(\frac{l}{L_P}\right)^2 H_0 t_c$. 

Our results show that the available limits on $\alpha$ variation are
inconsistent with the scale length of the theory $l$ being larger than
Planck scale.

\section{Summary and Discussion}
\label{conclusiones}

In this paper, we have analyzed the variation of $\alpha$ in the early
universe. We have modified the Kawano code, CAMB and CosmoMC in order
to include the possible variation of $\alpha$. We have used recent
observational abundances of light elements to obtain bounds on
$\frac{\Delta \alpha}{\alpha_0}$ at the time of primordial
nucleosynthesis. We have used recent data from the CMB and the 2dFGRS
power spectrum to limit the variation of $\alpha$ at recombination.
Results obtained in sections \ref{nucleo} and \ref{cmb} are consistent
with no variation of $\alpha$ during primordial nucleosynthesis and
recombination of neutral hydrogen. 

It is important to check that the values of the baryon density
obtained using the light elements abundances (section \ref{nucleo})
are consistent with the respective value obtained using data from the
CMB (section \ref{cmb}). Using the relation $\eta_B = 2.739 \times
10^{-8} \Omega_B h^2$, we find that results are consistent within
$1\sigma$. 

We have also used our results from the early universe and
recent bounds from the late universe to test Bekenstein model.  
We have improved the analysis of the Bekenstein model with respect to a
previous work \citep{LV02} in various aspects: i) we have obtained
analytical expressions for the Bekenstein model which include the
dependence on $t_c$ (while other authors put $t_c=0$) for the
variation of $\alpha$ in two regimes: a) radiation and matter and
b) matter and cosmological constant, ii) the whole data set is updated.

On the other hand, E\"otv\"os-like experiments provide stringent
constraints on the Bekenstein model parameters. Constraints for
$\frac{l}{L_P}$ can be set using these kind of experiments (see
appendix \ref{eotvos}). We obtain  $\frac{l}{L_P}<8.7 \times 10^{-3}$
which is one order of magnitude below the limits obtained in this
paper using astronomical and geophysical data: $\frac{l}{L_P}<6 \times
10^{-2}$. Nevertheless, the importance of our analysis lies on the
fact that while E\"otv\"os-like experiments test planetary scales, in
this paper we test different time scales, namely cosmological
time-scales. 

The values obtained for the free parameters of the model
disagree with the supposition that $l>L_p$, implied in Bekenstein's
framework. However, this latter requirement could be relaxed. Indeed,
it should be noted that Bekenstein's framework is 
very similar to the dilatonic sector of string theory, and it has been
pointed out that in the context of string theories
\citep{strings1,strings2} there is no need for an universal relation
between the Planck and the string scale.

\appendix
\section{E\"otv\"os-like experiments}
\label{eotvos}
A general expression
for the E\"otv\"os parameter and recent calculations on the
proton-neutron mass difference were performed by \citet{CV02}. From
this work it follows that: 
\begin{equation}
\frac{\Delta a}{a}=\Gamma_E \Delta \left(\frac{E_{em}}{Mc^2}\right)
\end{equation}
where $a$ is the acceleration for different bodies falling freely in a
gravitational field $g$, $E_{em}$ is the electromagnetic energy
of the falling bodies and $M$ is the nucleon mass at rest. A bound on
$\Gamma_E$ was estimated in the same paper: $\left| \Gamma_E
\right|<1.2 \times 10^{-9}$. Comparing this expression with equation
(45) of \citet{Bekenstein82}:
\begin{equation}
\frac{\Delta a}{a}=  \frac{1}{2\pi} \zeta \left(\frac{l}{L_P}\right)^2
\Delta\left( \frac{E_{em}}{Mc^2}\right)
\end{equation}
it follows that: 
\begin{equation}
\Gamma_E = \frac{1}{2\pi} \zeta \left(\frac{l}{L_P}\right)^2
\end{equation}
We obtain $\frac{l}{L_P}<8.7 \times 10^{-3}$.

\section*{{\bf Acknowledgements}}

Support for this work was provided by Project G11/G071, UNLP and PIP
5284 CONICET.  The authors would like to thank Andrea Barral, Federico
Bareilles, Alberto Camyayi and Juan Veliz for technical and
computational support. The authors would also like to thank Ariel
Sanchez for support with CosmoMC. MEM wants to thank Sergio Iguri for
the helpful discussions. CGS gives special thanks to Licia Verde and
Nelson Padilla for useful discussion. SJL wants to thank Michael
Murphy for useful discussions. The authors are grateful to Jacob
Bekenstein for valuable advice.

\bibliography{bibliografia3} 
\bibliographystyle{aa}

\end{document}